# Journal of Economic Literature codes classification system (JEL)

Jussi T. S. Heikkilä[1]

*This version: July 2022*

[See ISKO Encyclopedia of Knowledge Organization version at https://www.isko.org/cyclo/jel]

Abstract

The Journal of Economic Literature codes classification system (JEL) published by the *American Economic Association* (AEA) is the *de facto* standard classification system for research literature in economics. The JEL classification system is used to classify articles, dissertations, books, book reviews, and working papers in EconLit, a database maintained by the AEA. Over time, it has evolved and extended to a system with over 850 subclasses. This paper reviews the history and development of the JEL classification system, describes the current version, and provides a selective overview of its uses and applications in research. The JEL codes classification system has been adopted by several publishers, and their instructions are reviewed. There are interesting avenues for future research as the JEL classification system has been surprisingly little used in existing bibliometric and scientometric research as well as in library classification systems.

[1] Jyväskylä University School of Business and Economics, PO Box 35, FI-40014 University of Jyväskylä, Finland. Lappeenranta-Lahti University of Technology LUT, LUT School of Engineering Science, Lahti, Finland. email: jussi.heikkila@jyu.fi

# 1 Introduction

Journal of Economic Literature codes classification system (JEL) is a domain-specific hierarchical alphanumeric classification scheme—and a core knowledge organization system (Mazzocchi 2018)—in the field of economics. The *Journal of Economic Literature*[1] itself (also abbreviated JEL; for the sake of clarity, we use the abbreviation only for the JEL code system) is a journal published by the *American Economic Association* (AEA) since 1969, and its mission is to help keep AEA members informed of research developments in various fields of economics. The JEL codes classification system is "well-established in economics and most of the papers published in economics journals have JEL codes attached" (Bornmann and Wohlrabe 2019, p. 843). Hence, it is the *de facto* standard method of classifying scholarly literature in the field of economics (cf. Ekwurzel 1995; Pencavel 2008; Kempf and Neubert 2016; Cherrier 2017).

This article aims to provide an overview of the current knowledge related to the JEL codes classification system. The rest of the article is structured as follows. Section 2 reviews the origins and history of the JEL code system. Section 3 discusses the coverage and structure of the JEL codes classification system. Section 4.1 examines the JEL system in relation to library classifications, 4.2 considers the use of the system and review instructions by major publishers and selected economic journals, and 4.3 considers online repositories' use of JEL codes. Section 5 discusses selected applications of the JEL code system in research, and Section 6 concludes.

# 2 Origins and history of the JEL code system[2]

The current version of the JEL classification system (EconLit subject descriptors) is accessible online on the webpage of the AEA.[3] As the JEL classification system of today is a result of long-term development by the AEA, it is important to briefly describe the history that led to the creation of the JEL codes classification system. The AEA was established on September 9, 1885, at a meeting in Saratoga, New York, when there was concurrently the second meeting of the *American Historical Association* (Ely 1936).[4] Since the first meeting with less than 50 participants (Ely 1936), the association has, as of 2022 and according to AEA's webpage, grown to more than 20,000 members from academe, business, government, and consulting groups who are professionals or graduate-level students dedicated to economics research and teaching.[5] It is a scholarly society dedicated to the discussion and publication of economics research.

In 1911, the AEA founded the *American Economic Review* (AER) as a journal for its members (Coats 1969; Ekwurzel 1995). During the first half of the 20[th] century, the development of classification codes in the field of economics seems to not have been particularly systematic, although there were increasingly classification initiatives, and, for instance, a directory for listed AEA members by self-reported "fields of interests" was developed (AEA 1948; Cherrier 2017). Cherrier (2017) identifies four major revisions to the AEA's classification system, which were undertaken in 1938–44, 1955–56, 1966, and 1988–90. According to this source, the first efforts by AEA members to classify economic literature and personnel were influenced by the Second World War as government agencies of the United States needed to draft economists into the war and rebuilding efforts. It is therefore important to keep in mind that at the time, the need for economics classification was also driven by the requirement to register economists' fields of expertise in the United States through the *National Register of Scientific and Technical Personnel* and subsequently through the *National Science Foundation* (NSF), as noted by Cherrier (2017).

In the 1950s and 1960s, economic personnel and the literature were both proliferating (Cherrier 2017). As a partial solution to the burgeoning economics literature, in 1961, a committee of the AEA began to publish the *Index of Economic Journals* to facilitate access to the literature of economics (Ekwurzel 1995). In 1963, the AEA started publishing the *Journal of Economic Abstracts* (Coats 1969), and in 1966, this journal was included in members' AER subscriptions. Concurrently, a swelling number of economics publications, budgetary constraints of the AEA, and computerization (spearheaded by fields such as chemistry, biology, and medicine) led to the need to rationalize and automate the AEA's bibliographic efforts (Cherrier 2017). Finally, *the Journal of Economic Literature* replaced *the Journal of Economic Abstracts* in 1969.

The JEL codes classification system was originally developed for use in the *Journal of Economic Literature*. Prof. John Pencavel, the editor of the *Journal of Economic Literature 1986-1998*, declared in his Report of the Editor (1990, p.476) that "the mission of the *Journal of Economic Literature* is to help members of the Association keep abreast of research developments in various fields of economics. This goal is affected by providing a bibliographic guide to research publications, reviews of certain books, and articles describing and evaluating research progress on particular topics." As noted, the JEL system is used to classify articles, dissertations, books, book reviews, and working papers in EconLit (explained below, see Ekwurzel and Saffran 1985; Ekwurzel 1995; Millhorn 2000; Zhang and Su 2018) and in many other applications. In addition to the online version, the classification is published in the *Journal of Economic Literature*, which is published quarterly. Almost all major publishers have journals that use JEL classification codes (see Section 4.2). Currently, the quarterly issues of the *Journal of Economic Literature* include survey articles on economic literature, book reviews, an annotated index of new books in economics, a content listing and subject index of journal articles, and abstracts of articles from selected journals. [6]

The early JEL bibliography was stored on magnetic tapes, which made it possible to mount the *Economic Literature Index* (ELI) on the *DIALOG Information Retrieval Service* in 1981 (Ekwurzel 1995). [7] Ekwurzel and Saffran (1985) described searching techniques for this database, and regarding subject descriptors (see Appendix 2, pp.1757-1761), they noted (p. 1731) that they differ from JEL codes as they are one level more fine-grained: "all citations in the ELI carry four-digit subject descriptor codes. It is important for the researcher to understand that the ELI descriptor codes are the subject classifications used in the annual Index of Economic Articles and not the three-digit subject headings used in the JEL Subject Index of Articles in Current Periodicals. The four-digit subject descriptor codes are actually breakdowns of the JEL three-digit classifications. Each citation retrieved on the ELI displays all the subject descriptor codes (up to seven) assigned by JEL staff members to the article." We revisit the number of assigned JEL classification codes in the following.

In 1991, AEA published the ELI index in CD-ROM bibliography in partnership with *SilverPlatter Information*—one of the first companies to create reference databases on CD-ROM—and this new digital database was called EconLit (Ekwurzel 1995, Pencavel 2008). [8] Thus, originally, EconLit was the CD-ROM version of the ELI. The coverage of EconLit is discussed briefly in Section 3. The year 1995 marked an important milestone in the digitization of economic research as the first CD-ROM version of the full text of the *Journal of Economic Literature* issue was published, and this was, according to Pencavel (1996, p. 7), "the first journal in economics available in this form." There is less systematic information available on how the development of the online availability of EconLit evolved over time concurrently with the increasing availability of CD-ROMs. Ekwurzel (1995, p.105) notes that "by 1992, journal and abstract coverage in the online bibliography was expanded to include both journals and abstracts not indexed in JEL" and that more than 430 economics journals were indexed online in 1994. In the Internet era, the AEA web (i.e., AEA's homepage) was designed to be a one-stop portal for economists, where most of the content could be accessed, including online journals access for AEA members, and where also "[t]he JEL classification system, widely used by economics journals to classify articles by subject, can be reached with two clicks" (Ekwurzel and McMillan 2001, pp.7–8). It should also be noted that EconLit has been a significant source of revenue for the AEA (as "a monopoly of information about Economics literature"), and this revenue has been used to keep the prices of the AEA's flagship journals low and to launch new journals: the *Journal of Economic Perspectives* and AEA-sponsored domain-specific journals (Pencavel 2008, p.9, see Section 4.2).

Cherrier (2017) provides the most comprehensive historical description of the evolution of the JEL classification system. The first version of JEL codes was introduced in 1969, and the most recent version of main classes begins in 1991 (Cherrier 2017; Kosnik 2018). According to Cherrier (2017), in the past, incremental changes to the JEL classification system were decided upon and implemented through exchanges between the JEL board of editors and the Pittsburgh office, where the bibliographical department was managed (first by Naomi Perlman and from 1985 by Drucilla Ekwurzel). One can access the up-to-date classification and related JEL classification codes guide online.[9] Interestingly, very little research exists on the evolution of the JEL classification system itself until recently (Cherrier 2017; Kosnik 2018).

Table 1 presents the JEL classification systems and its evolution at the level of main classes. First, for comparison, there is the 1969 classification system that was published in the first issue of the *Journal of Economic Literature* (AEA 1969) and then the 1986 system (AEA 1990) just before a major change in 1991 (AEA 1991a, 1991b). The system on the right is also the current system. Cherrier (2017) vividly describes the evolution of the JEL classification system and discusses the underlying dynamics and debates in the Executive Committee of the AEA and between leading economists.[10] A significant change in the new 1991 version was the creation of Microeconomics and Macroeconomics categories that caused the disappearance of the previous "Theory" category, whose existence and content had been debated for half a century, according to Cherrier (2017). According to Prof. John Pencavel, who was leading the effort to create the new version of JEL codes in 1988-1990, the elimination of theory reflected his perspective that (1) the best theory papers are those that combine theory with empirical work and (2) that theory should be a component of all the other categories. [11]

**Table 1. Comparison between different classification systems, the highest level**

| 1969 JEL Classification System for Articles and Abstracts | 1990 JEL Classification System for Articles and Abstracts | 1991 JEL Classification System for Books and Articles | 2021 JEL Classification System |
|---|---|---|---|
| 000 General economics; Theory; History; Systems | 000 General Economic Theory; History; Systems | A General Economics and Teaching | A General Economics and Teaching |
| 100 Economic growth; Development; Planning; Fluctuations | 100 Economic Growth; Development; Planning; Fluctuations | B Methodology and History of Economic Thought | B History of Economic Thought, Methodology, and Heterodox Approaches |
| 200 Economic Statistics | 200 Quantitative Economic Methods and Data | C Mathematical and Quantitative Methods | C Mathematical and Quantitative Methods |
| 300 Monetary and fiscal theory and institutions | 300 Domestic Monetary and Fiscal Theory and Institutions | D Microeconomics | D Microeconomics |
| 400 International Economics | 400 International Economics | E Macroeconomics and Monetary Economics | E Macroeconomics and Monetary Economics |
| 500 Administration; Business finance; Marketing; Accounting | 500 Administration; Business Finance; Marketing; Accounting | F International Economics | F International Economics |
| 600 Industrial organization; Technological change; Industry studies | 600 Industrial Organization; Technological Change; Industry Studies | G Financial Economics | G Financial Economics |
| 700 Agriculture; Natural resources | 700 Agriculture; Natural Resources | H Public Economics | H Public Economics |
| 800 Manpower; Labor; Population | 800 Manpower; Labor; Population | I Health, Education, and Welfare | I Health, Education, and Welfare |
| 900 Welfare programs; Consumer economics; Urban and regional economics | 900 Welfare Programs; Consumer Economics; Urban and Regional Economics | J Labor and Demographic Economics | J Labor and Demographic Economics |
| | | K Law and Economics | K Law and Economics |
| | | L Industrial Organizations | L Industrial Organizations |
| | | M Business Administration and Business Economics; Marketing; Accounting | M Business Administration and Business Economics; Marketing; Accounting; Personnel Economics |
| | | N Economic History | N Economic History |
| | | O Economic Development, Technological Change, and Growth | O Economic Development, Innovation, Technological Change, and Growth |
| | | P Economic Systems | P Economic Systems |
| | | Q Agriculture and Natural Resource Economics | Q Agriculture and Natural Resource Economics; Environmental and Ecological Economics |
| | | R Urban, Rural, and Regional Economics | R Urban, Rural, and Regional, Real Estate, and Transportation Economics |
| | | Z Miscellaneous | Y Miscellaneous Categories |
| | | | Z Other Special Topics |

Notes: AEA (1969) provides the original JEL classification. AEA (1990) and AEA (1991a) classifications illustrate the big change in 1991. AEA (1991b) provides full mapping between the old pre-1991 and new 1991 JEL classifications. AEA (2021) presents the most recent version. See also Cherrier (2017) for a broader history of the changes and details.

According to prof. Pencavel, when he as the editor of the *Journal of Economic Literature* initiated and was assigned the task to revise the JEL classification (Pencavel 2008), he suggested the first level of classification by letters and then asked members of his JEL board to supply the next level of classification by numbers. The idea of this two-level classification and revision was to facilitate search efforts, and more generally, there was a need to respond both to the expanding volume of research in the well-established areas of economics and to research in economics invading less conventional areas of social science.[12] Cherrier (2017, footnote 51, p. 574) provides information on the specialists in charge of developing the structure of each category. Most of them were members of the JEL board of editors at the time. Many other economists also commented on the draft structures before Pencavel, Drucilla Ekwurzel, and Asatoshi Maeshiro made them consistent and finalized the new classification. Table 2 lists these specialists.

**Table 2. Specialists (not an exhaustive list) involved in creating the 1991 JEL classification system**

| Proposed general categories in 1990* | Old classification system* | Specialists** (affiliations in 1990) |
|---|---|---|
| A General Economics | 010, 200 | John J. Siegfried (Vanderbilt University) |
| B Methodology and History of Economic Thought | 030 | John K. Whitaker (University of Virginia) |
| C Mathematical and Quantitative Methods | 200 | Asatoshi Maeshiro (University of Pittsburgh) |
| D Microeconomics | 021, 022, 024, 025, 026, 902 | John H. Pencavel (Stanford University) |
| E Macroeconomics | 023, 130 | Alan S. Blinder (Princeton University) |
| F International Economics | 400 | Richard C. Marston (University of Pennsylvania) |
| G Financial Economics | 310 | Thomas Mayer (University of California, Davis) |
| H Public Economics | 320 | Harvey S. Rosen (Princeton University) |
| I Health, Education, and Welfare | 910 | Robert A. Moffit (Brown University) |
| J Labor and Demographic Economics | 800 | John Pencavel (Stanford University) |
| | | Mark R. Killingsworth (Rutgers University) |
| K Law and Economics | 916 | A. Mitchell Polinsky (Stanford University) |
| | | Steven M. Shavell (Harvard University) |
| L Industrial Organization | 610, 630, 640 | Roger G. Noll (Stanford University) |
| M Business Economics and Accounting | 500 | |
| N Economic History | 040 | Moses Abramovitz (Stanford University) |
| | | Alexander J. Field (Santa Clara University) |
| O Economic Development, Technological Change, and Growth | 110, 121, 122, 620 | Glenn L. Nelson (University of Minnesota) |
| P Economic Systems | 027, 050, 113, 123, 124 | Duncan K. Foley (Columbia University) |
| Q Agriculture and Natural Resource Economics | 700 | Daniel A. Sumner (North Carolina State University) |
| R Urban, Rural, and Regional Economics | 930, 940 | Edwin S. Mills (Northwestern University) |
| Z Miscellaneous | | |

Notes: *Based on Pencavel (1990, Table 1, p. 477) **Based on Cherrier (2017, footnote 51, p. 574) and email to author from Prof. Pencavel. As noted by Cherrier, many other economists commented on the drafts before Pencavel, Ekwurzel, and Maeshiro finalized the classification.

Cherrier (2017) notes that the AEA was contemplating a major revision to the JEL codes in the meeting of the executive committee in 2013; however, Table 3 suggests that the changes to the JEL codes classification system between 2013 and 2022 have remained rather incremental thus far. A few categories have been removed, while much more new codes have been added, including E7 Macro-Based Behavioral Economics, G4 Behavioral Finance, G5 Household Finance, Z2 Sports Economics, and Z3 Tourism Economics with respective 3-digit subclasses. In addition, the titles of dozens of JEL codes have been amended. In the era of increasingly better capabilities to perform quantitative and qualitative analyses with full-text data of articles and topic modelling, there seems to be a decreasing need for author- and editor-assigned classifications, at least in the case of search techniques. Cherrier (2017 p.569) also explains that due to the development of ELI/EconLit, the "literature search was increasingly done through keywords and less through JEL code filtering, although the latter was still dominant in the 1980s."

**Table 3. Incremental changes between 2013 and 2022**

| Deleted JEL codes | | New JEL codes | |
|---|---|---|---|
| D03 | Behavioral Microeconomics: Underlying Principles | B17 | History of Economic Thought through 1925: International Trade and Finance |
| | | | K16 Election Law |
| D92 | Interpersonal Firm Choice: Investment, Capacity, and Financing | B27 | History of Economic Thought since 1925: International Trade and Finance |
| | | | K24 Cyber Law |
| D99 | Intertemporal Choice: Other | B55 | Social Economics |
| E03 | Behavioral Macroeconomics | | K25 Real Estate Law |
| | | D15 | Intertemporal Household Choice; Life Cycle Models and Saving |
| | | | K38 Human Rights Law; Gender Law |
| G02 | Behavioral Finance: Underlying Principles | D16 | Collaborative Consumption |
| | | | O35 Social Innovation |
| | | D25 | Intertemporal Firm Choice: Investment, Capacity, and Financing |
| | | | O36 Open Innovation |
| | | D26 | Crowd-Based Firms |
| | | | P18 Energy; Environment |
| | | E14 | Austrian; Evolutionary; Institutional |
| | | | Q35 Hydrocarbon Resources |
| | | E70 | Macro-Based Behavioral Economics: General |
| | | | Z20 Sports Economics: General |
| | | E71 | Macro-Based Behavioral Economics: Role and Effects of Psychological, Emotional, Social, and Cognitive Factors on the Macro Economy |
| | | | Z21 Sports Economics: Industry Studies |
| | | F45 | Macroeconomic Issues of Monetary Unions |
| | | | Z22 Sports Economics: Labor Issues |
| | | G40 | Behavioral Finance: General |
| | | | Z23 Sports Economics: Finance |
| | | G41 | Behavioral Finance: Role and Effects of Psychological, Emotional, Social, and Cognitive Factors on Decision Making in Financial Markets |
| | | | Z28 Sports Economics: Policy |
| | | G50 | Household Finance: General |
| | | | Z29 Sports Economics: Other |
| | | G51 | Household Saving, Borrowing, Debt, and Wealth |
| | | | Z30 Tourism Economics: General |
| | | G52 | Household Finance: Insurance |
| | | | Z31 Tourism: Industry Studies |
| | | G53 | Household Finance: Financial Literacy |
| | | | Z32 Tourism and Development |
| | | G59 | Household Finance: Other |
| | | | Z33 Tourism: Marketing and Finance |
| | | H13 | Economics of Eminent Domain; Expropriation; Nationalization |
| | | | Z38 Tourism: Policy |
| | | I26 | Returns to Education |
| | | | Z39 Tourism: Other |
| | | K15 | Civil Law; Common Law |

| Changed titles of JEL codes* | | |
|---|---|---|
| | 2013 | 2022 |
| A11 | Role of Economics; Role of Economists | Role of Economics; Role of Economists; **Market for Economists** |
| B13 | History of Economic Thought: Neoclassical through 1925 (Austrian, Marshallian, Walrasian, Stockholm School) | History of Economic Thought: Neoclassical through 1925 (Austrian, Marshallian, Walrasian, **Wicksellian**) |
| B16 | History of Economic Thought: Quantitative and Mathematical | History of Economic Thought **through 1925:** Quantitative and Mathematical |
| B25 | History of Economic Thought since 1925: Historical; Institutional; Evolutionary; Austrian | History of Economic Thought since 1925: Historical; Institutional; Evolutionary; Austrian; **Stockholm School** |
| B52 | Current Heterodox Approaches: Institutional; Evolutionary | Current Heterodox Approaches: **Historical;** Institutional; Evolutionary; **Modern Monetary Theory** |
| C24 | Single Equation Models; Single Variables: Truncated and Censored Models; Switching Regression Models | Single Equation Models; Single Variables: Truncated and Censored Models; Switching Regression Models; **Threshold Regression Models** |
| C25 | Single Equation Models; Single Variables: Discrete Regression and Qualitative Choice Models; Discrete Regressors; Proportions | Single Equation Models; Single Variables: Discrete Regression and Qualitative Choice Models; Discrete Regressors; Proportions; **Probabilities** |
| C32 | Multiple or Simultaneous Equation Models: Time-Series Models; Dynamic Quantile Regressions; Dynamic Treatment Effect Models; Diffusion Processes | Multiple or Simultaneous Equation Models: Time-Series Models; Dynamic Quantile Regressions; Dynamic Treatment Effect Models; Diffusion Processes; **State Space Models** |
| C55 | Modeling with Large Data Sets | Large Data Sets: Modeling **and Analysis** |
| C57 | Econometrics of Games | Econometrics of Games **and Auctions** |
| D02 | Institutions: Design, Formation, and Operations | Institutions: Design, Formation, Operations, **and Impact** |
| D04 | Microeconomic Policy: Formulation; Implementation; Evaluation | Microeconomic Policy: Formulation, Implementation, and Evaluation |
| D40 | Market Structure and Pricing: General | Market Structure, Pricing, and **Design:** General |
| D41 | Market Structure and Pricing: Perfect Competition | Market Structure, Pricing, and **Design:** Perfect Competition |
| D42 | Market Structure and Pricing: Monopoly | Market Structure, Pricing, and **Design:** Monopoly |
| D43 | Market Structure and Pricing: Oligopoly and Other Forms of Market Imperfection | Market Structure, Pricing, and **Design:** Oligopoly and Other Forms of Market Imperfection |
| D64 | Altruism; Philanthropy | Altruism; Philanthropy; **Intergenerational Transfers** |
| D74 | Conflict; Conflict Resolution; Alliances | Conflict; Conflict Resolution; Alliances; **Revolutions** |
| D83 | Search; Learning; Information and Knowledge; Communication; Belief | Search; Learning; Information and Knowledge; Communication; Belief; **Unawareness** |
| D90 | Intertemporal Choice: General | **Micro-Based Behavioral Economics:** General |
| D91 | Intertemporal Household Choice; Life Cycle Models and Saving | **Micro-Based Behavioral Economics: Role and Effects of Psychological, Emotional, Social, and Cognitive Factors on Decision Making** |
| E11 | General Aggregative Models: Marxian; Sraffian; Institutional; Evolutionary | General Aggregative Models: Marxian; Sraffian; **Kaleckian** |
| E12 | General Aggregative Models: Keynes; Keynesian; Post-Keynesian | General Aggregative Models: Keynes; Keynesian; Post-Keynesian; **Modern Monetary Theory** |
| E22 | Capital; Investment; Capacity | Investment; **Capital; Intangible Capital;** Capacity |
| E24 | Employment; Unemployment; Wages; Intergenerational Income Distribution; Aggregate Human Capital | Employment; Unemployment; Wages; Intergenerational Income Distribution; Aggregate Human Capital; **Aggregate Labor Productivity** |
| F02 | International Economic Order | International Economic Order **and Integration** |
| G10 | General Financial Markets: General (includes Measurement and Data) | **Asset Markets and Pricing** |
| G12 | Asset Pricing; Trading Volume; Bond Interest Rates | **Equities; Fixed Income Securities** |
| H26 | Tax Evasion | Tax Evasion **and Avoidance** |
| I12 | Health Production | Health **Behavior** |
| K32 | Environmental, Health, and Safety Law | **Energy,** Environmental, Health, and Safety Law |
| L31 | Nonprofit Institutions; NGOs | Nonprofit Institutions; NGOs; **Social Entrepreneurship** |
| L62 | Automobiles; Other Transportation Equipment | Automobiles; Other Transportation Equipment; **Related Parts and Equipment** |
| L65 | Chemicals; Rubber; Drugs; Biotechnology | Chemicals; Rubber; Drugs; Biotechnology; **Plastic** |
| L67 | Other Consumer Nondurables | Other Consumer Nondurables: **Clothing, Textiles, Shoes, and Leather Goods; Household Goods; Sports Equipment** |
| M00 | Business Administration and Business Economics; Marketing; Accounting: General | Business Administration and Business Economics; Marketing; Accounting; **Personnel Economics:** General |
| O00 | Economic Development, Technological Change, and Growth | Economic Development, **Innovation,** Technological Change, and Growth |
| O30 | Technological Change; Research and Development; Intellectual Property Rights: General | **Innovation;** Research and Development; Technological Change; Intellectual Property Rights: General |
| O32 | Management of Technological Innovation and R&D; | Management of Technological Innovation and R&D |
| O47 | Measurement of Economic Growth; Aggregate Productivity; Cross-Country Output Convergence | **Empirical Studies of Economic Growth;** Aggregate Productivity; Cross-Country Output Convergence |
| Q02 | Global Commodity Markets | Commodity Markets |
| Q52 | Pollution Control Adoption Costs; Distributional Effects; Employment Effects | Pollution Control Adoption **and** Costs; Distributional Effects; Employment Effects |
| Q54 | Climate; Natural Disasters; Global Warming | Climate; Natural Disasters **and Their Management;** Global Warming |
| R41 | Transportation: Demand, Supply, and Congestion; Safety and Accidents; Transportation Noise | Transportation: Demand, Supply, and Congestion; **Travel Time;** Safety and Accidents; Transportation Noise |
| Z13 | Economic Sociology; Economic Anthropology; Social and Economic Stratification | Economic Sociology; Economic Anthropology; **Language;** Social and Economic Stratification |

# 3 Coverage and structure

According to the AEA's official webpage of the JEL, "the JEL classification system was developed for use in the *Journal of Economic Literature* (JEL) and is a standard method of classifying scholarly literature in the field of economics. The system is used to classify articles, dissertations, books, book reviews, and working papers in EconLit and in many other applications." [13] As described, EconLit is a specialized database for economics literature and is maintained by the AEA (Ekwurzel and Saffran 1985; Ekwurzel 1995; Millhorn 2000; Zhang and Su 2018; Rose 2021). According to the AEA's official webpage of EconLit:

> "Professionally classified, updated weekly, and including over 1.6 million records, EconLit covers economics literature published over the last 130 years from leading institutions in 74 countries. In combination with the optional full-text package of over 500 journals, including the prestigious AEA journals, EconLit provides a comprehensive library of economics literature." [14]

Ekwurzel (1995) describes the coverage of EconLit and the *Economic Literature Index* during their earlier development phases. Recently, Rose (2021) has compared EconLit and Scopus as part of his replication study and reports that EconLit has larger coverage (in Economics and neighboring fields), whereas Scopus has more sophisticated author disambiguation. Table 1 illustrates how the main classes of the JEL code system have evolved over time, and Table 4 presents the current number of different-level JEL codes. Interestingly, there are significant differences in how fine-grained the two- and three-digit subcategories are across 20 main classes.

**Table 4. The number of JEL codes**

| 2021 JEL Classification System | Number of two-digit JEL codes | Number of three-digit JEL codes |
|---|:---:|:---:|
| A General Economics and Teaching | 3 | 16 |
| B History of Economic Thought, Methodology, and Heterodox Approaches | 5 | 32 |
| C Mathematical and Quantitative Methods | 9 | 70 |
| D Microeconomics | 9 | 65 |
| E Macroeconomics and Monetary Economics | 7 | 47 |
| F International Economics | 6 | 53 |
| G Financial Economics | 5 | 33 |
| H Public Economics | 8 | 56 |
| I Health, Education, and Welfare | 3 | 23 |
| J Labor and Demographic Economics | 8 | 62 |
| K Law and Economics | 4 | 30 |
| L Industrial Organizations | 9 | 72 |
| M Business Administration and Business Economics; Marketing; Accounting; Personnel Economics | 5 | 29 |
| N Economic History | 9 | 74 |
| O Economic Development, Innovation, Technological Change, and Growth | 5 | 41 |
| P Economic Systems | 5 | 43 |
| Q Agriculture and Natural Resource Economics; Environmental and Ecological Economics | 5 | 49 |
| R Urban, Rural, and Regional, Real Estate, and Transportation Economics | 5 | 32 |
| Y Miscellaneous Categories | 9 | 11 |
| Z Other Special Topics | 3 | 19 |
| Min | 3 | 11 |
| Max | 9 | 74 |
| Mean | 6.1 | 42.85 |
| Median | 5 | 42 |
| Total sum | 122 | 857 |

Notes: The source is the AEA: JEL Classification System / EconLit Subject Descriptors https://www.aeaweb.org/EconLit/jelCodes.php Accessed on 28 Feb 2022. "General" JEL codes under the highest-level codes with no 2nd level codes (e.g., B00, C00, C02, etc.) have been counted as three-digit JEL codes, as in Heikkilä (2021).

Figure 1 illustrates the notation in the case of one selected JEL class, "A14 Sociology of Economics." For each classification level, there are both guidelines and keywords; however, keywords are often not specified in the case of higher levels of JEL codes. At the most detailed level, there are also caveats mentioned and links to example articles given. Notably, the Guidelines and Caveats sections may include instructions regarding the choice of other JEL codes (e.g., cross-classifications) and information on the focal JEL codes in relation to other JEL codes, as demonstrated in Figure 1.

**Figure 1. Example of JEL code sub-class A14 Sociology of Economics**

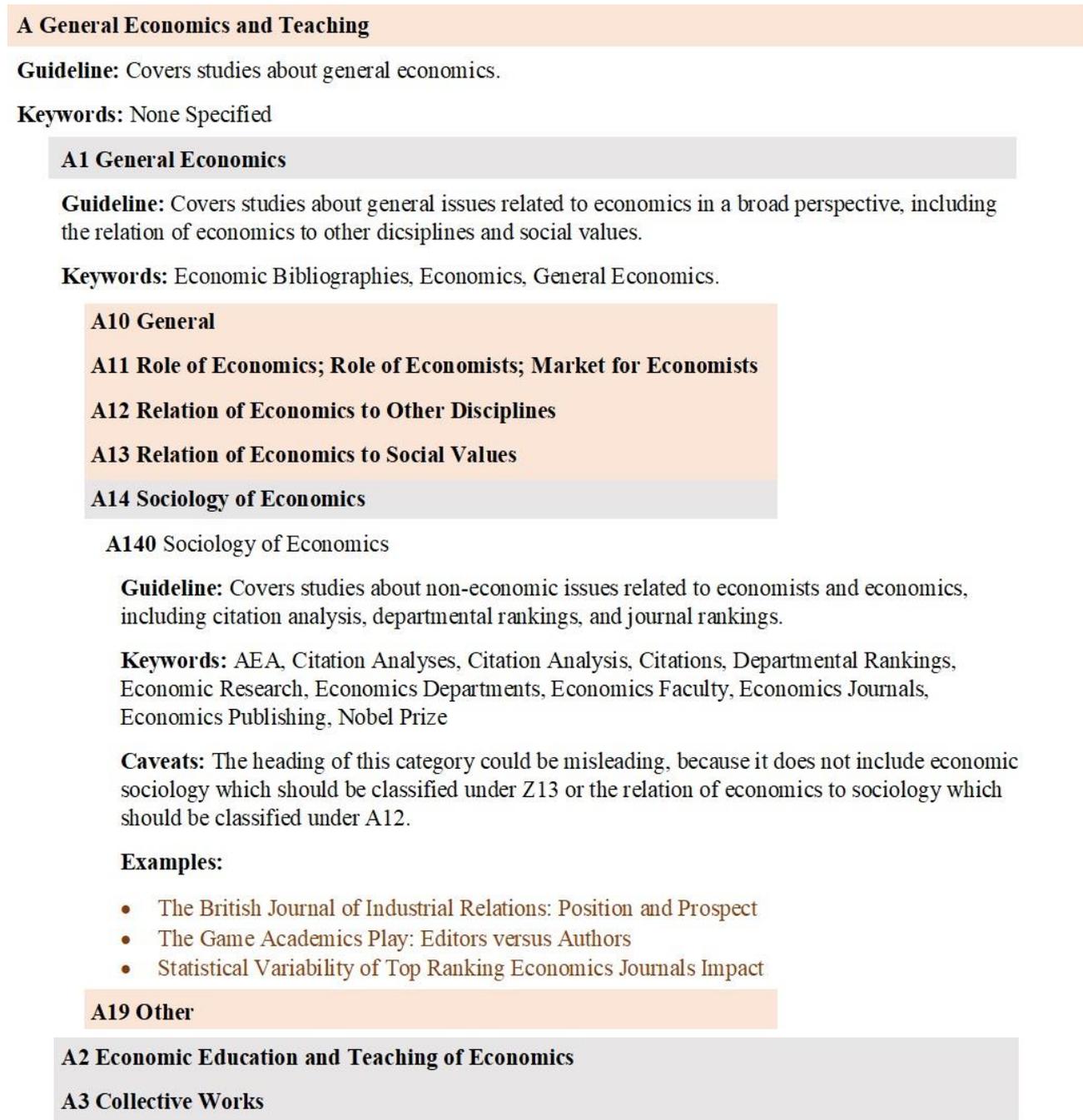

**A General Economics and Teaching**

**Guideline:** Covers studies about general economics.

**Keywords:** None Specified

**A1 General Economics**

**Guideline:** Covers studies about general issues related to economics in a broad perspective, including the relation of economics to other dicsiplines and social values.

**Keywords:** Economic Bibliographies, Economics, General Economics.

**A10 General**

**A11 Role of Economics; Role of Economists; Market for Economists**

**A12 Relation of Economics to Other Disciplines**

**A13 Relation of Economics to Social Values**

**A14 Sociology of Economics**

**A140** Sociology of Economics

**Guideline:** Covers studies about non-economic issues related to economists and economics, including citation analysis, departmental rankings, and journal rankings.

**Keywords:** AEA, Citation Analyses, Citation Analysis, Citations, Departmental Rankings, Economic Research, Economics Departments, Economics Faculty, Economics Journals, Economics Publishing, Nobel Prize

**Caveats:** The heading of this category could be misleading, because it does not include economic sociology which should be classified under Z13 or the relation of economics to sociology which should be classified under A12.

**Examples:**

- The British Journal of Industrial Relations: Position and Prospect
- The Game Academics Play: Editors versus Authors
- Statistical Variability of Top Ranking Economics Journals Impact

**A19 Other**

**A2 Economic Education and Teaching of Economics**

**A3 Collective Works**

Notes: Author's illustration based on the AEA's JEL Classification Codes Guide, https://www.aeaweb.org/jel/guide/jel.php

# 4 The use of the JEL code system

The different document types classified by the JEL code system have already been listed, and Figure 2 illustrates the historical number of documents classified using JEL codes by document type, according to Ekwurzel (1995). Journal articles comprise the majority of indexed documents; however, working papers are also crucial in the field of economics as publication lags are long (Ellison 2000, 2002; Lusher et al. 2021), and as of 2021, the average time from submission to a journal to its acceptance exceeds two years (Lusher et al. 2021).

According to Smith (2008), since 2007, *EBSCO Publishing* has been offering an "EconLit with Full Text" database that at the time of the launch incorporated full-text articles for 401 of the more than 750 economic journals indexed in EconLit. [15] The number of indexed journals has evolved over time, and as of February 2022, there were more than 2015 distinct *International Standard Serial Number* (ISSN) entries in the list of journals indexed in EconLit. [16] It should be noted that some of these journals have discontinued or changed names. Of the 2015 journals, EconLit reportedly has coverage to the present day for some three quarters (1455) of the distinct journals.

**Figure 2. Types of documents classified using JEL 1969-1994 according to Ekwurzel (1995)**

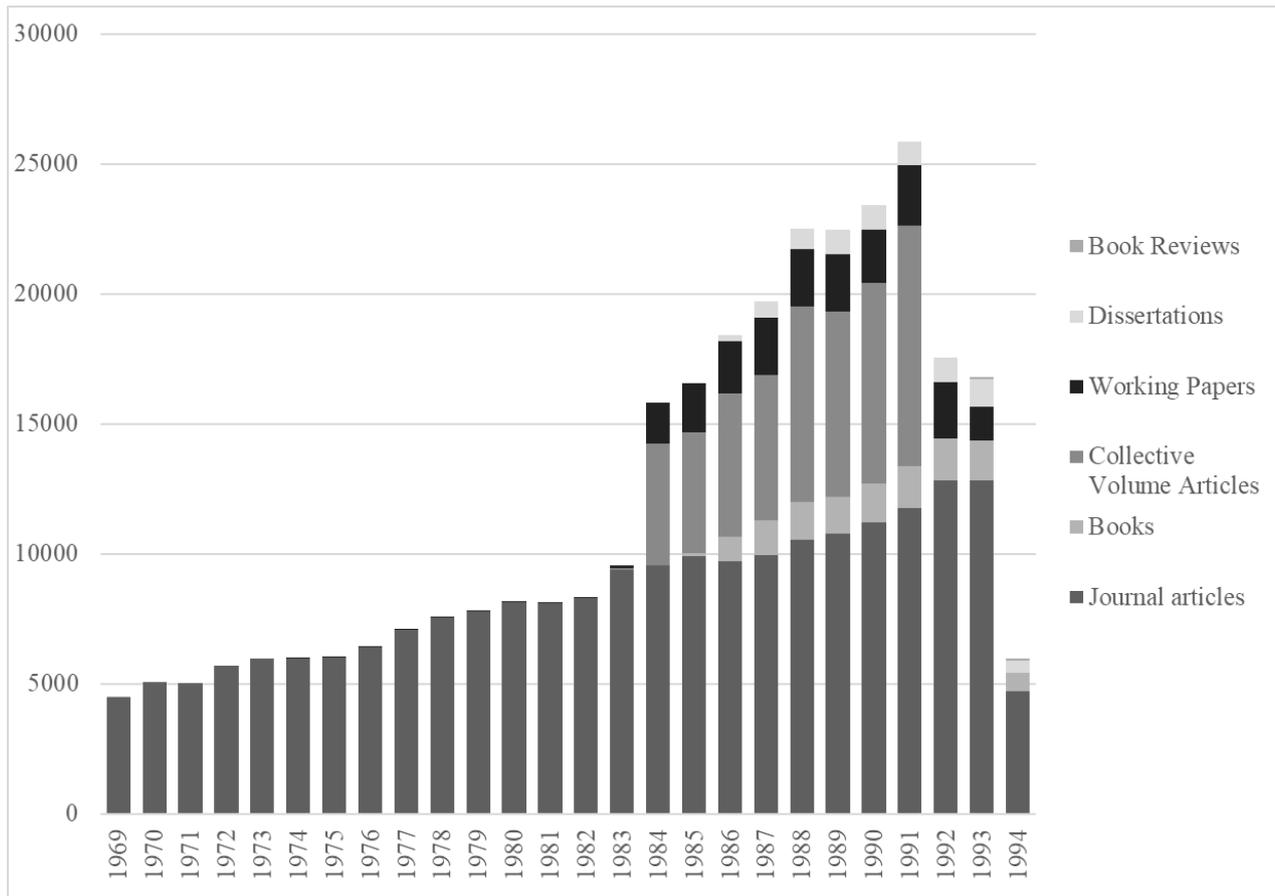

Notes: Source is Ekwurzel (1995, Table 1 on p. 106). Information for the years 1992-1994 is incomplete.

Pencavel (1991, v), the Editor of the *Journal of Economic Literature* at the time, noted that when introducing the new classification system in 1991, "members of the Association will appreciate that the allocation of an article or a book to a particular category is difficult, if not arbitrary, when the subject matter is relevant to more than one topic. This is a problem with any Classification System of this sort. We intend to continue our current practice of listing articles and books under one or at most two categories where appropriate. (In the other versions of our data base, our on-line information retrieval system and our recent CD-ROM version of our bibliographic data, articles are also listed under up to six cross classifications.)." This illustrates the differing classification approaches by the type of document: books were assigned one to two JEL codes while articles up to six.

EconLit assigns up to six three-digit JEL classification codes to publications (Boppart and Staub 2016, more on the number of JEL codes in the next section). Krueger (1999) reports the percentage of articles published by 1-digit JEL codes in the leading economics journals (AER, *Journal of*

*Economic Literature*, *Quarterly Journal of Economics* (QJE), *Journal of Political Economy* (JPE), *Journal of Economic Perspectives* (JEP)) and members of the AEA between January 1994 and July 1998. As some authors have reported more than one JEL code for their articles, Krueger (1999) used the first JEL code listed to classify the articles in these cases. Krueger's (1999) statistics show that Class D Microeconomics was the most common class, followed by E Macroeconomics and Monetary Economics. Clearly, some classes were more popular than others. D, E, F, G, H, I, J, L, and O each had a 5% or larger share of author-assigned classes, whereas B, C, K, M, N, P, Q, and R each had less than a 5% share of author-assigned JEL codes. Similarly, Boschini and Sjögren (2007) report that these listed nine most frequent JEL codes each account for more than 5% of the 4040 articles published in AER, JPE, and QJE between 1991 and 2002. Categories A General Economics and Teaching and Z Other Special Topics are peculiar categories that are less research-focused (also excluded in the analysis by Kosnik [2018]).

It should be noted that the majority of analyses focusing on the evolution of fields in economics by JEL codes has focused on a specific group of top journals. In the field of economics, there is clear domination by top journals (Ellison 2002; Kim et al. 2006b; Card and Della Vigna 2013; Hamermesh 2013; Linnemer and Visser 2016; Heckman and Moktan 2020). The often cited and analyzed "top five" journals are the *American Economic Review*, *Econometrica*, the *Journal of Political Economy*, the *Quarterly Journal of Economics*, and the *Review of Economic Studies*. Heckman and Moktan (2020) critically note that economics faculties' hiring, promotion, tenure, and prize committee discussions assess candidates based on the number of top five articles they have published or have in the pipeline and the rapidity with which they were generated. In addition, research proposals are often appraised by their potential to generate publication in the top five journals.

## 4.1 The use of JEL codes in libraries and library classifications

To our knowledge, there are no data available or analyses conducted on whether and to what extent libraries have been using JEL codes to organize library collections; however, one can search for anecdotal evidence from the webpages of university libraries, where search techniques for finding articles are described. There are some cases in which JEL codes are mentioned or where the browsing of articles by JEL code is made possible. [17] At the website of the University of Cambridge, one can search by JEL code for *Cambridge Working Papers in Economics* (CWPE). [18] As the JEL codes are naturally economics-specific, they cannot be used as a general library classification system. Nevertheless, one should expect library classifications to have used it for part of their general classification that is concerned with economic literature, but just a few hints of such uses have been found.[19]

Libraries use different library classification systems, including *Dewey Decimal Classification* (DDC), *Universal Decimal Classification* (UDC), *Library of Congress Classification* (LCC), and *Bliss Bibliographic Classification* (BC2), or other systems, including their own individual classifications (for more library systems, see Petrova and Petrov [2017] and the *ISKO Encyclopedia of Knowledge Organization*). Table 5 compares the main economics-related subclasses of these systems to JEL codes.

Recently, Petrova and Petrov (2017) reviewed economics classifications in different library classification systems but did not compare them to the JEL classification system "due to its widespread 'by default' in economic literature" (p.267). The authors concluded that "the libraries around the world are using different classification schemes, and in these schemes the role and place of economic science varies. There is no unity about what the main fields of economic science should be. In some classification schemes, for one area are given priority, while in other classification schemes the same areas are not used. This lack of a unified classification of economic sciences hinders the definition of the scope of economic science" (Petrova & Petrov 2017, p.267).

To our knowledge, it has not been documented whether the development of the JEL classification codes system has had an impact or has been impacted by other library classification systems; however, Cherrier (2017) notes that in the 1950s, librarians from the *Library of Congress* urged the AEA to provide a new classification scheme that could serve as a reference point for all the institutions dealing with economic literature. More recently, a semi-automatic mapping between JEL codes and *STW (Standard-Thesaurus Wirtschaf) Thesaurus for Economics* published by the *ZBW-Leibniz Information Centre for Economics* has been developed (Kempf and Neubert 2016; Rebholz et al. 2016) and can be accessed online. [20] Table 5 includes the main classes of the *STW Thesaurus for Economics* for comparison.

**Table 5. Comparison of the main classes of selected library classifications and JEL codes**

| | Library classifications | | | | The Standard-Thesaurus Wirtschaft (STW) Thesaurus for Economics*** | JEL codes |
| | Dewey Decimal Classification (DDC) | Universal Decimal Classification (UDC) | Library of Congress Classification (LCC)* | Bliss Bibliographic Classification (BC2)** | | |
|---|---|---|---|---|---|---|
| **Broader concept / Superclass** | 300 Social sciences | 3 Social Sciences | H Social Sciences | | | |
| **Class** | 330 Economics | 33 Economics. Economic science | HB Economic theory. Demography | T Economics, policital economy | V Economics | |
| **Narrower concepts / Subclasses** | 330 Economics | 330 Economics in general | HB1-3840 Economic theory. Demography | TB Economic history, economic conditions, descriptive economics | V.00 Economics | A General Economics and Teaching |
| | 331 Labor economics | 331 Labour. Employment. Work. Labour economics. Organization of labour | HB71-74 Economics as a science. Relation to other subjects | TBI Economic processes | V.01 Economic theory and methodology | B History of Economic Thought, Methodology, and Heterodox Approaches |
| | 332 Financial economics | 332 Regional economics. Territorial economics. Land economics. Housing economics | HB75-130 History of economics. History of economic theory Including special economic schools | TJL Wealth & income | V.02 Microeconomics | C Mathematical and Quantitative Methods |
| | 333 Economics of land and energy | 334 Forms of organization and cooperation in the economy | HB131-147 Methodology | TK Economic resources | V.03 Macroeconomics | D Microeconomics |
| | 334 Cooperatives | 336 Finance | HB135-147 Mathematical economics. Quantitative methods. Including econometrics, input-output analysis, game theory | TMB Forms of production, industrial structure | V.04 Business cycles and growth, economic structure | E Macroeconomics and Monetary Economics |
| | 335 Socialism and related systems | 338 Economic situation. Economic policy. Management of the economy. Economic planning. Production. Services. Prices | HB201-206 Value. Utility | TME U Units of economic activity, level of aggregation | V.05 Money and financial markets | F International Economics |
| | 336 Public finance | 339 Trade. Commerce. International economic relations. World economy | HB221-236 Price | TMX P Economic systems | V.06 Economic systems | G Financial Economics |
| | 337 International economics | | HB238-251 Competition. Production. Wealth | TPY Economics of specific infustries & services | V.07 International economics | H Public Economics |
| | 338 Production | | HB501 Capital. Capitalism | | V.08 Development economics | I Health, Education, and Welfare |
| | 339 Macroeconomics and related topics | | HB522-715 Income. Factor shares | | V.09 Public finance | J Labor and Demographic Economics |
| | | | HB535-551 Interest | | V.10 Industrial organization | K Law and Economics |
| | | | HB601 Profit HB615-715 Entrepreneurship. Risk and uncertainty. Property | | V.11 Regional science V.12 Environmental and resource economics | L Industrial Organizations M Business Administration and Business Economics; Marketing; Accounting; Personnel Economics |
| | | | HB801-843 Consumption. Demand HB846-846.8 Welfare theory HB848-3697 Demography. Population. Vital events HB3711-3840 Business cycles. Economic fluctuations | | V.13 Labour V.14 Social economics, education and health economics V.15 Economic history | N Economic History O Economic Development, Innovation, Technological Change, and Growth P Economic Systems Q Agriculture and Natural Resource Economics; Environmental and Ecological Economics R Urban, Rural, and Regional, Real Estate, and Transportation Economics Y Miscellaneous Categories Z Other Special Topics |

Notes: *There are also other subclasses relevant to economics, e.g., HC Economic history and conditions (cf. Petrova and Petrov 2017). **Source: *The Bliss Classification Bulletin* (1986) ***Source: https://zbw.eu/stw/version/latest/thsys/v/about.en.html Accessed 17 Mar 2022.

## 4.2 JEL code instructions by publishers and top economics journals

The variety of journals published by the AEA has increased to nine as of 2022.[21] On the AEA maintained webpages of each of these journals, there are no separate JEL code instructions; however, the list of JEL codes is available both in JEL format (hierarchical) and EconLit format (non-hierarchical) on the webpage of the AEA under the section "EconLit," and there are some EconLit Search Hints that also include the use of JEL codes or a "Subject classification system." [22] Moreover, the "JEL Classification Codes Guide" is available on the AEA webpage under the section "Resources." [23]

Table 6 presents the instructions for JEL codes by top economics journals of selected major academic publishers. The selection of "top journals" was based on the IDEAS ranking of journals so that for each major publisher (see Larivière et al. 2015), a journal that was among the highest ranking according to IDEAS/RePEc (*Research Papers in Economics*) Aggregate Rankings for Journals was chosen. Most of the reviewed journals did not explicitly guide or instruct on the selection of JEL codes. Moreover, the instructions, if given, were at a relatively high level, and there were links to the JEL classification codes guide on the AEA webpage. In contrast to other publishers, *Oxford University Press* provides a separate filtering option for JEL codes on the webpage's navigation pane of each of their journals and general information on JEL codes. [24] It is noted that "JEL codes are submitted by the article's authors; the codes are displayed on the print article, usually after the abstract, and are used to build a 'Browse by JEL code' listing on the journal's website." [25] *Oxford University Press* also provides the possibility to sign up for JEL code alerting so that one receives an email when research related to a specific JEL code is published.

Taylor & Francis' webpage has a section that responds to the question, "Can I search for economics articles by JEL code?" [26] There is a positive answer: "You can search for new content in our economics journals using JEL (Journal of Economic Literature) codes. Most journals use these codes, allowing authors to tag individual articles with the most relevant JEL number(s). To search for papers by JEL code, simply type the JEL code into the search bar. For example, to search for all recent economics papers about 'D12 - Consumer Economics: Empirical Analysis,' type 'D12' into the search field. Then click the Search icon to return all recent articles that have been tagged with this particular JEL code."

**Table 6. Selected publishers' instructions**

| Publisher (alphabetical order) | Top ranked economics journal, example* | JEL code instructions** |
|---|---|---|
| American Economic Association | American Economic Review | The official JEL Classification Codes Guide on AEA webpage |
| Cambridge University Press | The Journal of Economic History | - |
| Chicago University Press | Journal of Political Economy | - |
| De Gruyter | The B.E. Journal of Macroeconomics | - |
| Econometric Society | Econometrica | - |
| Elsevier | Journal of Financial Economics | **Classification codes** Please provide up to 6 standard JEL codes. The available codes may be accessed at JEL. |
| Emerald | Journal of Economic Studies | - |
| Oxford University Press | The Quarterly Journal of Economics | **JEL codes** Authors should include at least 1 JEL code with manuscript during submission. (Authors may provide several, depending on the number of subject area the article encompasses.) JEL codes should be included at the end of the abstract. They also should be provided as a combination of one letter and two numbers (e.g., Q04). If you have any question regarding JEL codes, please visit this website. |
| SAGE | The American Economist | - |
| Springer | Journal of Economic Growth | **Classification code** JEL An appropriate number of JEL codes should be provided. This classification system is prepared and published by the Journal of Economic Literature, see https://www.aeaweb.org/econlit/jelCodes.php?view=jel |
| Taylor & Francis | Journal of Business & Economic Statistics | - |
| Wiley | Journal of Applied Econometrics | - |

Notes: Notes: *For each publisher, these journals are among the highest-ranking journals according to IDEAS/RePEc Aggregate Rankings for Journals. Available at: https://ideas.repec.org/top/top.journals.all.html **JEL code instructions were collected from the webpages (Submission instructions for authors sections) of journals in February 2022.

Caruso and Campiglio (2007) report that among top economic journals, the average number of JEL codes per article ranged between 1.2 in *The Quarterly Journal of Economics* and 3.3 in the *American Economic Review*. Observations by Boppart and Staub (2016) corroborate the large differences between journals. Moreover, they note that "while half the articles fall into exactly one field according to the one-digit definition, about 37 percent contribute to two fields, and somewhat over 10 percent have three one-digit JEL codes" (Boppart and Staub 2016, p.19). Interestingly, in top economics journals published by the AEA, authors assign their own JEL codes, but editors may then assign their own codes (Kosnik 2018).

Kosnik (2018) has analyzed this phenomenon and reports that there is indeed a statistically significant disparity in use of JEL code assignments between editors and authors for the same papers and that authors assign more JEL codes than editors. Kosnik (2018, p.253) reports that editors of the AER assigned between 1990-2008, on average, 2.57 JEL codes per full-length research paper, while authors assigned 2.73 codes. Kosnik (2018) further notes that while authors tend to assign more JEL codes to their papers, they are distinguished often by differing subcategories and not by broad categories compared to editors, and while editors assign fewer total JEL codes per paper, they seem to assign more codes to articles crossing discipline boundaries. Boppart and Staub (2016, Figuer A.1 in Appendix) illustrate that the average number of JEL codes per article has evolved over time: they show that the increase between 1991 and 2009 is clearly visible at the 1-digit, 2-digit, and 3-digit levels in 50 "core journals of economics." In 1991, the average number of 3-digit JEL codes was less than two, whereas in 2009, it was approaching three.

While Boppart and Staub (2016) note that final JEL codes can and do differ from JEL codes declared by authors, Kosnik (2018) focused on this phenomenon and systematically studied the differences between author and editor assigned JEL codes in the AER between 1990 and 2008. She reports that there were significant differences. [27] Table 7 provides a few examples of differences between author-assigned JEL codes in working papers and editor-assigned JEL codes in final published articles. For instance, the article by Cherrier (2017) published in the *Journal of Economic Literature* is assigned only one JEL code A14 "Sociology of Economics," which according to the JEL classification codes guide[28], "covers studies about non-economic issues related to economists and economics, including citation analysis, departmental rankings, and journal rankings" (see Figure 1). In a working paper version, Cherrier (2015) had listed five JEL codes—A10, A14, B10, B20, and B00—so, in this case, there was a significant reduction in the JEL codes. In line with Kosnik's (2018) observations, the authors in most of these examples assigned more JEL codes in their working paper versions compared to the final published version of the articles. On the other hand, there are also cases (e.g., Acemoglu et al. 2001; Rodrik et al. 2004) where the final published article is assigned more JEL codes than the working paper.

**Table 7. Differences between author-assigned and editor-assigned JEL codes, examples**

| Working paper version | Initial JEL codes (number) | Final version | Final JEL codes (number)* | Change(s) in JEL codes | Note |
|---|---|---|---|---|---|
| Cherrier (2015) | A10, A14, B10, B20, B00 (5) | Cherrier (2017) | A14 (1) | Decrease (-4) | |
| Kim et al. (2006a) | A11, B20, O33 (3) | Kim et al. (2006b) | A11, A14 (2) | Decrease (-1) | |
| Card & DellaVigna (2013a) | A1, A11 (2) | Card &DellaVigna (2013b) | A14 (1) | Decrease (-1) and change (1) | |
| Hamermesh (2012) | B20, J24 (2) | Hamermesh (2013) | A14 (1) | Decrease (-1) and change (1) | |
| Kosnik (2015) | A1, B0 (2) | Kosnik (2018) | A14, D85 (2)** | Change (2) | |
| Heckman & Moktan (2018) | A14, I23, J44, O31 (4) | Heckman & Moktan (2020) | A14, I23, J44, J62 (4) | Change (1) | |
| Bornmann & Wohlrabe (2017) | A110, A120, A140 (3) | Bornmann & Wohlrabe (2019) | A11, A12, A14 (3)** | No change | |
| Faria (2000) | A19, C79, L19 (5) | Faria (2005) | A19, C79, L19 (3) | No change | Article given as an example in the JEL codes guide for class A14 (see Figure 1) |
| Rodrik et al. (2002) | O1, O4 (2) | Rodrik et al. (2004) | F1, N7, O1 (3) | Increase (+1) and change (1) | JEL codes at 2-digit level |
| Acemoglu et al. (2000) | O11, P16, P51 (3) | Acemoglu et al. (2001) | O11, P51, I12, N10, O57 (5) | Increase (+2) and change (1) | One of the most cited papers in economics |
| Ellison (2000) | A14 (1) | Ellison (2002) | A11, A14 (2)** | Increase (+1) | |

Notes: *Final print version as the primary source of information **Information from EconLit

## 4.3 Online repositories

While the AEA's EconLit (Ekwurzel and Saffran 1985; Ekwurzel 1995; Millhorn 2000; Zhang & Su 2018) is probably the most authoritative source of economic literature, it is proprietary and requires a subscription. Here, we briefly review selected open-access repositories of economics research that use JEL codes. It should be noted that many online repositories, such as arXiv, do not provide the alternative to search by JEL codes as they are an economics-specific classification system. There are also discontinued repository initiatives in the field of economics, such as Economists Online (Puplett 2010).

According to its webpage, IDEAS/RePEc is the largest bibliographic database dedicated to economics, and it is freely available on the Internet, indexing 3.9 million items of research as of February 2022. [29] One can browse these research documents by JEL codes. JEL code information available in RePEc has been used in multiple studies (e.g., Rath and Wohlrabe 2016; Orazbayev 2017; Colussi 2018; Zacchia 2021).

Elsevier's SSRN (formerly Social Science Research Network) is a repository for preprints of research in social sciences, humanities, life sciences, health sciences, and more. In SSRN, the JEL codes field under the "Keywords" field is optional when submitting papers to SSRN. [30] One can browse research by these author-assigned JEL codes on SSRN.

The EconBiz portal is a service from *ZBW - Leibniz Information Centre for Economics* to search for economics and business research (Kempf et al. 2016), and EconStor[31] is a related publication server and online repository for economic research. While one can also view JEL codes of the search result documents if they are available in the original publication, there is no alternative to search by JEL code under the "Advanced" search option; however, a project has been launched to create a mapping between the *STW (Standard-Thesaurus Wirtschaft) Thesaurus for Economics* extended search terms and JEL classification codes (Kempf and Neubert 2016; Rebholz et al. 2016). [32]

*National Bureau of Economic Research* (NBER) working papers are among the most influential working papers in economics (Lusher et al. 2021), and authors may assign JEL codes when publishing their working papers in NBER working paper series. [33] According to the NBER webpage, more than 1200 non-peer-reviewed working papers are published each year by NBER affiliates, and papers issued more than 18 months previously are open-access. As of 27 Feb 2022, the number of the latest working paper was 29785; however, currently, one cannot search or browse working papers by JEL code on the NBER webpage as the "Topics" filter differs from the JEL classification. [34]

# 5 Applications of the JEL classification system in research
In addition to decreasing the search costs when searching for economic research, JEL codes enable multiple applications in research (see e.g., Kosnik 2018). Next, we briefly describe selected examples.

JEL codes have been used to determine sub-populations in empirical analyses of economics research. For instance, Corsi et al. (2010) distinguish between mainstream and heterodox (JEL codes B50, B51, B52, B53, B54, B59, E11, and E12) economics using JEL codes when analyzing pluralism among Italian economists. D'Orlando (2013) analyzes citation counts as a measure of scientific relevance in the five theoretical schools classified in JEL code B5 "Current Heterodox Approaches." Wagstaff and Culyer (2012) use JEL codes in their bibliometric analysis of health economics, while Fernandez et al. (2021) use an unsupervised machine-learning algorithm (Latent Dirichlet Allocation) to analyze economic education scholarly work (JEL code A2) and to identify "hidden" topics in the field. Claveau et al. (2021) use JEL codes in their analysis of the evolution of the philosophy of economics by particularly focusing on B4 Economic Methodology.

A particular research application is the evolution of economic research topics and fields based on JEL codes or some other classification system (e.g., AEA 1948; Perlman and Perlman 1977; Laband and Wells 1998; Kim et al. 2006b; Karbownik and Knauff 2009; Kelly and Bruestle 2011; Boppart and Staub 2016; Linnemer and Visser 2016) Here, it is important to distinguish between the evolution of the classification scheme itself and the evolution of the allocation of attention by economists across the JEL classes. Cherrier (2017, p. 545) suggests that "the history of the classification system used by the *American Economic Association* (AEA) to list economic literature and scholars is a relevant proxy to understand the transformation of economics science throughout the twentieth century." Davis (2019, p.275) argues that "the changes in the JEL code that Cherrier identifies are the product of an increasingly diverse research frontier in economics." Several studies have used different levels of JEL codes, or JEL codes have served as a basis to create tailored or aggregated categories for economics papers (e.g., Ellison 2002; Kim et al. 2006b; Card and Della Vigna 2013; Davis 2019; Lundberg and Stearns 2019; Angrist et al. 2020; Card et al. 2020).

An obvious challenge in such classification exercises is the fact that articles are often assigned multiple, typically 1 to 6, JEL codes (Caruso and Campiglio 2007; Boppart and Staub 2016), which leads to either some sort of simplification, such as classifying the documents according to the first JEL code, or double counting. For instance, Ellison (2002) used JEL codes in combination with rules based on title keywords and paper-by-paper judgements to assign a sample of economics papers into 17 fields. Kim et al. (2006b) identifies fields of highly cited economics papers by collecting the first JEL code listing in EconLit. Card and Della Vigna (2013) have created 14 mutually exclusive fields based on JEL codes and have classified each article "up to five fields based on the first five JEL codes in EconLit" (p. 157). Card and DellaVigna (2013) as well as Kelly and Bruestle (2011) treat an article with n different codes as n different articles, with each assigned a weight of 1/n for an article.

Researchers are increasingly using JEL codes to analyze diversity in economics (e.g., Caruso and Campiglio 2007) and differences between female and male economists, among other topics (e.g., Laband and Wells 1998; Lundberg and Stearns 2019; Card et al. 2020). For instance, Laband and Wells (1998) report that female economists are historically more likely to contribute scholarship to the areas of Labor Economics and Welfare programs, Consumer Economics, and Urban and Regional Economics (JEL codes 800 and 900 in the pre-1991 classification). More recently, Lundberg and Stearns (2019) analyzed the differences between male and female economists and reported that the distributions of men and women across seven economics fields based on the JEL codes of their doctoral dissertations are highly similar, but women's higher representation in labor and public fields is apparent. The data come from the *Doctoral Dissertations in Economics* lists published annually in the *Journal of Economic Literature*, which represents almost all major PhD-granting departments in the US over the period from 1991-2017 (Lundberg and Stearns 2019). JEL codes have also been used to analyze the evolution and diversity of research topics by country (e.g., Corsi et al. 2010, 2019, for Italy).

Boppart and Staub (2016) analyze how digitization and the online accessibility of economics journals is associated with "innovational strength of follow-on research" (p.1). They operationalize this measure of "innovational strength" by using an article's "number of unique JEL codes as a measure of the breadth of an article's content" (p. 19) and analyze the impact of online accessibility. They document that online accessibility has led to more innovative follow-up research according to this innovational strength measure. Recently, several studies have reported network analyses and network visualizations of connections between JEL codes (e.g., Kosnik 2018; Bickley et al. 2021; Larrosa 2021).

There are also various other applications. Laband and Wells (1998) report that the length of articles published in AER, JPE, and QJE varies by JEL codes and note that among other things, articles written on general economic theory (JEL subject code 000, pre-1991 JEL classification) are significantly shorter compared to most others. Nowell and Grijalva (2011) explain that co-authorship appears to differ by JEL code: the share of single-authored papers is highest in N Economic History and B History of Thought, whereas G Financial Economics and Q Agricultural and Natural Resource are the most prone to co-authorship. JEL codes have also been used to control for research-field-level fixed effects (such as domain-specific citation patterns) in regression analyses (e.g., Axarloglou and Theoharakis 2003; Boschini and Sjögren 2007; Card and DellaVigna 2013; Boppart and Staub 2016; Bornmann and Wohlrabe 2019). Recently, Heikkilä (2021) attempted to link the JEL codes to the *United Nation's Sustainable Development Goals* (SDGs) using keywords and shows that each SDG has some corresponding JEL classes based on keyword overlap. As a practical application, the JEL classification system is used to classify job postings in the *American Economic Association's* Job Openings for Economists (JOE) Listings online service.[35]

## 6 Conclusion

The JEL codes classification system, maintained by the AEA, has remained the *de facto* standard classification system for economic literature since 1969. There are currently no competing classification systems in the field of economics, with the exception of some library classifications. The latest big change to the system was implemented in 1991, and since then, there have been incremental changes and occasional additions to the system (Cherrier 2017). The JEL classification system has grown to more than 850 sub-categories, which the AEA's JEL codes guide helps to use. While initially, economics classifications in the United States were influenced by the need to classify economists and draft them into the war effort and into rebuilding the country during and after the second World War, the current JEL classification system is specifically used to classify articles, dissertations, books, book reviews, and working papers in EconLit, a specialized database for economics literature maintained by the AEA.

We are not aware of a large-scale or systematic use of JEL codes by libraries, and there is little evidence of the JEL codes' effects on major library classification systems; however, some online repositories (e.g., IDEAS/RePEc and SSRN) allow authors to browse articles by JEL codes, while some (e.g., arXiv) do not. We reviewed instructions for authors by major publishers regarding JEL codes and documented differences: some have instructions, while others do not use JEL codes at all. Generally, the journals that allow authors to assign JEL codes seem to have only very high-level instructions regarding how to choose JEL codes and no motivation on why it is important.

JEL codes have served as a basis for higher-level classifications (e.g., Ellison 2002; Card and Della Vigna 2013) and also to create JEL-code-level fixed effects to control for domain-specific citation patterns (Bornmann and Wohlrabe 2019). Researchers are increasingly using JEL codes to analyze, among other things, diversity in economics (Caruso and Campiglio 2007), and for instance, differences between female and male economists (Lundberg and Stearns 2019). There are several interesting avenues for future research. One interesting topic is to analyze the process regarding how authors choose the JEL codes in practice and to identify their motives. Are there differences between economists and non-economists?

Finally, digitalization and the Internet have also had major impacts on the importance of classification in the field of economics (Ekwurzel and McMillan 2001). Full-text searches in search engines have become increasingly available (Hjørland 2012), and the size of academic search engines, some of which are open-access, has grown tremendously (Gusenbauer 2019). Concurrently, as there is increasingly more computation power available, various machine-learning and natural language processing applications can be used to cluster documents based on the actual text instead of or in addition to author-assigned classification codes (e.g., Angrist et al. 2020; Fernandex et al. 2021). As noted by Fernandez et al. (2021, p.156), "the common use of JEL codes only identifies the academic setting for each paper but does not identify the underlying economic concept the paper addresses." Relatedly, Hamermesh (2013, p. 168) notes that while it is easy to obtain authors' classifications of their published papers by subject (JEL code), the subject does not automatically imply method because, for instance, field experiment methods can be used in diverse areas, such as industrial organization, labor economics, and public economics.

Szostak (2003) provided a simple way to classify theory types and methods based on five questions: "Who?", "What?", "Where?", "When?", and "Why?". Current and further developments of supervised and unsupervised topic modeling techniques and machine-learning tools can increasingly answer such questions. Hence, the importance of author-assigned JEL classification codes could be decreasing relative to automatic machine-augmented classification. How can the economic research community further develop and potentially automate the process of JEL code classification? This is another interesting topic for future research.

As stated by Hull (1998, p. 272), "the fundamental elements of any classification are its theoretical commitments, basic units and the criteria for ordering these basic units into a classification." Concerning the basic units of JEL-codes, they are the specialties and disciplines of economics, but what are the theoretical commitments and criteria for ordering these units into a classification? Cherrier (2017) provides a fine review of controversies in the history of JEL, of the heated discussions on the status of theoretical and empirical work, data, and measurement, and proper objects of analysis in economics as well as the influence of the contradictory demands of users, including economists, civil servants, journalists, publishers, librarians, and the military, and reflects on rapidly changing institutional and technological constraints. Cherrier (2017) also states that recent transformations were fueled by institutional and technical transformations rather than intellectual ones; however, a classification always reflects a view of what it classifies and is always serving some purposes better than others. Therefore, JEL is partaking in the theoretical struggles on the future development of the economic domain, and intellectual problems will never be superfluous. There is therefore a need for further research on the theoretical commitments of classification in economics.

**Acknowledgements**

The author would like to thank Birger Hjørland, John Pencavel, Beatrice Cherrier and two anonymous reviewers for their helpful comments and suggestions. Financial support from the Finnish Cultural Foundation (Päijät-Häme Regional Fund, Anja and Jalo Paananen Fund) and the Foundation for Economic Education (Liikesivistysrahasto, Reino Rossi Memorial Fund) is gratefully acknowledged.

Notes

1.  AEA: About *the Journal of Economic Literature*, https://www.aeaweb.org/journals/jel/about-jel Accessed 28 Feb 2022.
2.  We rely here heavily on Cherrier's (2017) extensive description of the history of JEL classification system. Also Ekwurzel (1995) provides an in-detail description of the practicalities related to the development of JEL classification system.
3.  AEA: JEL Classification System / EconLit Subject Descriptors, https://www.aeaweb.org/econlit/jelCodes.php Accessed 20 Feb 2022.
4.  Richard T. Ely (1854-1943) was the initiator, one of the founders and the first secretary of the AEA (Taylor 1944).
5.  AEA: About the AEA, https://www.aeaweb.org/about-aea Accessed 20 Feb 2022.
6.  AEA: About *the Journal of Economic Literature*, https://www.aeaweb.org/journals/jel/about-jel Accessed 28 Feb 2022.
7.  Cherrier (2017, p.568) reports that this happened in 1983.